\pdfoutput=1

\documentclass[reprint,amsmath,amssymb,floatfix,longbibliography,aip,letterpaper,cha,author-year]{revtex4-1}

\usepackage{graphics,graphicx,float}
\usepackage{mathtools}		

\makeatletter
\preprintsty@sw{}{%
 \usepackage{titlesec}
 \titleformat{\section}{\normalfont\footnotesize\sffamily\bfseries\uppercase}%
 	{\thesection}{1em}{}%
 \titleformat{\subsubsection}{\normalfont\small\sffamily\slshape}{\thesubsubsection}{1em}{}%
 }%
\makeatother
  
\def\autocite{\citep}

\usepackage[pdftex,
            pdfauthor={Steven A. Frank},
            pdftitle={The Inductive Theory of Natural Selection},
            pdfsubject={The Inductive Theory of Natural Selection}]{hyperref} 

\newcommand{\zbar}{\bar{z}}

\newcommand{\Rbar}{\bar{R}}
\newcommand{\gbar}{\bar{g}}

\newcommand{\cov}{\mathrm{Cov}}
\newcommand{\var}[1]{{V}_{#1}}

\newcommand{\R}{\mathcal R}

\newcommand{\GDs}{\GD_s}
\newcommand{\GDn}{\GD_n}
\newcommand{\GDg}{\GD_g}
\newcommand{\GDt}{\GD_t}
\newcommand{\GDc}{\GD_c}

\newcommand{\GDS}{\GD S}
\newcommand{\GDT}{\GD\Gt}
\newcommand{\subsub}[1]{_{_{#1}}}
\newcommand{\ssA}{\subsub{A}}
\newcommand{\ssW}{\subsub{W}}

\newcommand{\varss}[1]{{V}\subsub{#1}}
\newcommand{\preg}[2]{\Gb_{#1\cdot#2}}

\newcommand*{\Ga}{\alpha}
\newcommand*{\Gb}{\beta}
\newcommand*{\Gd}{\delta}
\newcommand*{\GD}{\Delta}
\newcommand*{\Ge}{\epsilon}
\newcommand*{\Gg}{\gamma}

\newcommand*{\Gm}{\mu}

\newcommand*{\Gt}{\tau}

\newcommand*{\dd}{\mathrm{d}}

\newcommand*{\Eq}[1]{Eq.~(\ref{eq:#1})}
\newcommand*{\Eqp}[1]{(Eq.~\ref{eq:#1})}

\newcommand*{\ovr}[2]{{\frac{#1}{#2}}}
\newcommand*{\dovr}[2]{\ovr{\dd #1}{\dd #2}}

\newcommand*{\prt}{\partial}
\newcommand*{\povr}[2]{\ovr{\prt #1}{\prt #2}}

\newcommand*{\Fig}[1]{Fig.~\ref{fig:#1}}

\newcount\BoxNum \BoxNum 1\relax
\makeatletter
\newcommand*{\boxlabel}[1]{%
  \protected@write \@auxout {}{\string \newlabel {box:#1}{{\the\BoxNum}}{}}%
  \advance\BoxNum 1\relax}
\makeatother

\newcommand*{\boldrule}{\hrule height 1.2pt}
\newcommand*{\noterule}{\medskip\boldrule\medskip}	

\renewcommand*{\Eq}[1]{Eq.~(\ref{eq:#1})}

\usepackage{titlesec}
  \titleformat{\section}[runin]{\normalfont\bf}{}{0pt}{}[.\ ]%

\renewcommand{\thesection}{\arabic{section}}
\renewcommand{\thesubsection}{\thesection.\arabic{subsection}}
\renewcommand{\thesubsubsection}{\thesubsection.\arabic{subsubsection}}

\begin{document}

\title{The Inductive Theory of Natural Selection: Summary and Synthesis}

\author{Steven A.\ Frank}
\affiliation{Department of Ecology and Evolutionary Biology, University of California, Irvine, CA 92697--2525  USA}

\begin{abstract}

The theory of natural selection has two forms. Deductive theory describes how populations change over time. One starts with an initial population and some rules for change. From those assumptions, one calculates the future state of the population. Deductive theory predicts how populations adapt to environmental challenge. Inductive theory describes the causes of change in populations. One starts with a given amount of change. One then assigns different parts of the total change to particular causes. Inductive theory analyzes alternative causal models for how populations have adapted to environmental challenge. This chapter emphasizes the inductive analysis of cause\footnote{homepage: \href{https://stevefrank.org}{https://stevefrank.org}}$^,$\footnote{A condensed and simplified version of this manuscript will appear as: Frank, S. A. and Fox, G. A. 2017. The inductive theory of natural selection. Pages 000--000 in \textit{The Theory of Evolution}, S.~M.\ Scheiner and D.~P.\ Mindell, eds. University of Chicago Press (in press).}.

\bigskip

\end{abstract}

\maketitle

{\renewcommand{\tocname}{}\small\hbox{\null}\vskip-41pt\tableofcontents}

\section{Introduction}

Darwin got essentially everything right about natural selection, adaptation, and biological design. But he was wrong about the processes that determine inheritance.

Why could Darwin be wrong about heredity and genetics, but be right about everything else? Because the essence of natural selection is trial and error learning. Try some different approaches for a problem. Dump the ones that fail and favor the ones that work best. Add some new approaches. Run another test. Keep doing that.  The solutions will improve over time. Almost everything that Darwin wanted to know about adaptation and biological design depended only on understanding, in a general way, how the traits of individuals evolve by trial and error to fit more closely to the physical and social challenges of reproduction.

Certainly, understanding the basis of heredity is important. Darwin missed key problems, such as genomic conflict. And he was not right about every detail of adaptation. But he did go from the absence of understanding to a nearly complete explanation for biological design. What he missed or got wrong requires only minor adjustments to his framework. That is a lot to accomplish in one step.

How could Darwin achieve so much? His single greatest insight was that a simple explanation could tie everything together. His explanation was natural selection in the context of descent with modification. Of course, not every detail of life can be explained by those simple principles. But Darwin took the stance that, when major patterns of nature could not be explained by selection and descent with modification, it was a failure on his part to see clearly, and he had to work harder. No one else in Darwin's time dared to think that all of the great complexity of life could arise from such simple natural processes. Not even Wallace.

Now, more than 150 years after \textit{The Origin of Species,} we still struggle to understand the varied complexity of natural selection. What is the best way to study the theory of natural selection: detailed genetic models or simple phenotypic models? Are there general truths about natural selection that apply universally? What is the role of natural selection relative to other evolutionary processes?

Despite the apparent simplicity of natural selection, controversy remains intense. Controversy almost always reflects the different kinds of questions that various people ask and the different kinds of answers that various people accept as explanations. Natural selection itself remains as simple as Darwin understood it to be.

\begin{table*}[!t]
  \caption{Natural Selection}
  \label{tab:natsel}
  \vskip2pt\noterule
  \noindent Domain: Evolutionary change in response to natural selection. 
  \noterule\vskip2pt
  \noindent Propositions:
  \begin{enumerate}\parskip=2pt\baselineskip=15pt\vskip-4pt
\item Evolutionary change can be partitioned into natural selection and transmission.
\item Adaptation arises as natural selection accumulates information about the environment.
\item Information is often lost during transmission of characters from ancestors to descendants.
\item The balance between information gain by selection and information loss by transmission often explains the relative roles of different evolutionary forces.
\item Fitness describes the evolutionary change by natural selection. 
\item Fitness can be partitioned into distinct causes, such as the amount of change caused by different characters.
\item Characters can be partitioned into distinct causes, such as different genetic, social, or environmental components. 
\item The theory of natural selection evaluates alternative causal decompositions of fitness and characters.
\item Key theories of natural selection identify sufficient causes of change. The fundamental theorem identifies the variance in fitness as a sufficient statistic for change by selection. Kin selection decomposes change by selection into sufficient social forces.    
\end{enumerate}
  \noterule
\end{table*}

\section{Constraints on selection}

One can look at the diverse and complicated biological world, and marvel at how much can be understood by the simple process of natural selection. Or one can look at the same world and feel a bit outraged at how much a naively simple-minded view misses of the actual complexity. That opposition between simplicity and complexity arose early in the history of the subject. 

\begin{figure}
\centering
\includegraphics[width=3.5in]{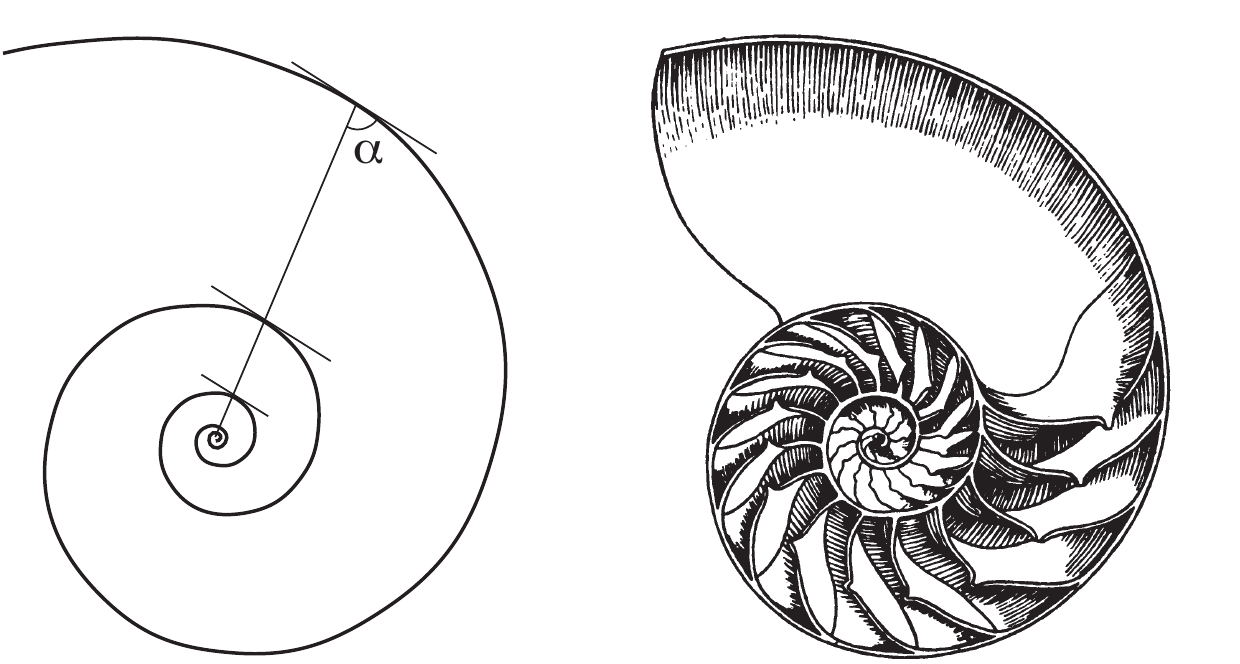}
\caption{The logarithmic spiral. The curve grows out at a rate that increases with the angle, $\Ga$, at the leading edge. The \textit{Nautilus} shell on the right closely follows a logarithmic spiral. The shell drawing on the right is from \textcite[p.~173]{thompson61on-growth}.}
\label{fig:logSpiral}
\end{figure}

D'Arcy Thompson (\citeyear{thompson17on-growth}) emphasized that physical processes influence growth and set the contours that biological design must follow. Those physically imposed contours limit natural selection as an explanation for organismal form. For example, mollusks often grow their shell by uniform addition at the leading edge. That uniform growth produces shells that follow a logarithmic spiral (\Fig{logSpiral}). 

A smaller angle of deposition at the leading edge causes tighter coiling of the shell. The physical laws of growth set the primary contours of pattern. Natural selection can modulate design only within those strongly constrained contours. Much of the order in nature arises from the physics of growth rather than from selection. 

Thompson applied that logic to a vast range of natural history. He showed that the great variety of shell patterns arises from just a few additional rules of growth. Natural selection apparently modulates only a small number of angles and rates of deposition. Similarly, the bizarrely diverse shapes of sheep, ram, and goat horns can be reduced to modulation of a few simple rules of growth. In general, a small number of generative processes in development set tightly constrained contours on the possible range of final form:
\begin{quote}
The distribution of forces which manifest themselves in the growth and configuration of a horn is no simple nor merely superficial matter.  One thing is co-ordinated with another; the direction of the axis of the horn, the form of its sectional boundary, the specific rates of growth in the mean spiral and at various parts of its periphery---all these play their parts, controlled in turn by the supply of nutriment which the character of the adjacent tissues and the distribution of the blood-vessels combine to determine.  To suppose that this or that size or shape of horn has been produced or altered, acquired or lost, \textit{by Natural Selection,} whensoever one type rather than another proved serviceable for defence or attack or any other purpose, is an hypothesis harder to define and to substantiate than some imagine it to be \autocite[p.~213]{thompson92on-growth}.
\end{quote} 
The rules of growth determine the range of forms that may occur. Physical processes constrain variation.

\section{The origin of variation}

The tension between the constraints on variation and the power of natural selection to shape observed pattern recurs throughout the history of post-Darwinian biology. If selection is trial and error, progress depends on the way in which new alternative trials arise. A trait cannot be selected if it never occurs. The origin of variation as a constraint on selection forms perhaps the greatest criticism against selection by itself as a creative force. Haldane (1932, p 94) framed the problem more broadly than Thompson:
\begin{quote}
It is perfectly true, as critics of Darwinism never tire of pointing out, that $\ldots$ no new character appears in the species as the result of selection. Novelty is only brought about by selection as the result of the combination of previously rare characters.
\end{quote}
How do rare characters first arise? That question is beyond the scope of a chapter on the theory of natural selection, because the origin of characters mostly depends on forces other than selection. But it is essential to keep this point in mind when considering the history of the theory of selection and the status of modern debates about how one should think about selection in relation to evolution.

\section{Selection by itself}

In another great book, Fisher agreed with Thompson about the complexity of evolutionary forces. Yet \textcite[p.~vii]{fisher30the-genetical} began his book by saying
\begin{quote}
Natural Selection is not Evolution.  Yet, ever since the two words have been in common use, the theory of Natural Selection has been employed as a convenient abbreviation for the theory of Evolution by means of Natural Selection, put forward by Darwin and Wallace. This has had the unfortunate consequence that the theory of Natural Selection itself has scarcely ever, if ever, received separate consideration. $\ldots$ The overwhelming importance of evolution to the biological sciences partly explains why the theory of Natural Selection should have been so fully identified with its role as an evolutionary agency, as to have suffered neglect as an independent principle worthy of scientific study. $\ldots$ The present book, with all the limitations of a first attempt, is at least an attempt to consider the theory of Natural Selection on its own merits.
\end{quote}
Natural selection does not stand alone. But selection remains the only evolutionary force that could potentially explain adaptation. With the warnings about the origin of variation in mind, I now turn to study of selection by itself.

\section{Goals of selection theory}

Selection theory must be evaluated with regard to two alternative goals. First, how can one improve predictions about evolutionary change in traits. Prediction is valuable when using artificial selection to enhance performance. For example, one may seek greater milk production from cows or reduced antibiotic resistance or stronger binding of a molecule to a cellular surface receptor. In these cases, the primary goal is improved prediction of outcomes to achieve greatest performance at least cost. Understanding the causes of the outcomes can be helpful and interesting, but causal analysis is secondary to performance. 

The second goal concerns the causal analysis of traits. How have various evolutionary forces shaped traits? Why do particular patterns of traits occur? Understanding cause depends on comparing the predictions of alternative explanatory models. However, prediction to evaluate cause differs from prediction to optimize costs and benefits with regard to a desired target. A causal model seeks to isolate the forces ultimately responsible for pattern, whereas a model to optimize performance may not provide a correct interpretation of cause. In this chapter, I focus on how to use selection theory to understand cause. 

\section{Partitioning causes of change}

\begin{quote}
Since path analysis depends on structure, and structure in turn depends on the cause-and-effect relationship among the variables, we shall first say a few words about the way these terms will be used. $\ldots$ There are a number of formal definitions as to what constitutes a cause and what an effect. For instance, one may think that a cause must be doing something to lead to something else (effect). While this is clearly one type of cause-and-effect relationship, we shall not limit ourselves to that type only. Nor shall we enter into philosophical discussions about the nature of cause-and-effect. We shall simply use the words `cause' and `effect' as statistical terms similar to independent and dependent variables, or [predictor variables and response variables] \autocite[p.~3]{li75path}.
\end{quote}

One can often partition total evolutionary change into separate components. Those separate components may sometimes be thought of as the separate causes of evolutionary change. The meaning of ``cause'' is of course a difficult problem. We are constrained by the fact that we only have access to empirical correlations, and correlation is not causation. Within that constraint, I follow Li's suggestion to learn what we can about causation by studying the possible structural relations between variables. Those structural relations express hypotheses about cause. Alternative structural relations may fit the data more or less well. Those alternatives may also suggest testable predictions that can differentiate between the relative likelihood of the different causal hypotheses \autocite{crespi90measuring,frank97the-price,frank98foundations,scheiner00using}. 

In evolutionary studies, one typically tries to explain how environmental and biological factors influence characters. Causal analysis separates into two steps. How do alternative character values influence fitness? How much of the character values is transmitted to following generations? These two steps are roughly the causes of selection and the causes of transmission. 

\section{Models of selection: prelude}

Improvement by trial and error is a very simple concept. But applying that simple concept to real problems can be surprisingly subtle and difficult. Mathematics can help, but it can also hinder. One must be clear about what one wants from the mathematics and the limitations of what mathematics can do. By \textit{mathematics,} I simply mean the steps by which one starts with particular assumptions and then derives logical conclusions or empirical predictions.  

The output of mathematics reflects only what one puts in. If different mathematical approaches lead to different conclusions, that means that the approaches have made different assumptions. Strangely, false or apparently meaningless assumptions often provide a better description of the empirical structure of the world than precise and apparently true assumptions. From \textcite[p.~ix]{fisher30the-genetical}
\begin{quote}
The ordinary mathematical procedure in dealing with any actual problem is, after abstracting what are believed to be the essential elements of the problem, to consider it as one of a system of possibilities infinitely wider than the actual, the essential relations of which may be apprehended by generalized reasoning, and subsumed in general formulae, which may be applied at will to any particular case considered. Even the word possibilities in this statement unduly limits the scope of the practical procedures in which he is trained; for he is early made familiar with the advantages of imaginary solutions, and can most readily think of a wave, or an alternating current, in terms of the square root of minus one.
\end{quote}
The immense power of mathematical insight from false or apparently meaningless assumptions shapes nearly every aspect of our modern lives. The problem with the intuitively attractive precise and realistic assumptions is that they typically provide exactness about a reality that does not exist. One never has a full set of true assumptions. By contrast, false or apparently meaningless assumptions, properly chosen, can provide profound insight into the logical and the empirical structure of nature. That truth may not be easy to grasp. But experience has shown it to be so over and over again. 

\section{Frequency change}

I begin with a basic model of fitness and frequency change. There are $n$ different types of individuals.  The frequency of each type is $q_i$.  Each type has $R_i$ offspring, where $R$ expresses reproductive success.  Average reproductive success is $\Rbar = \sum q_iR_i$, summing over all of the different types indexed by the $i$ subscripts. Fitness is $w_i=R_i/\Rbar$, used here as a measure of relative success. The frequency of each type after selection is 
\begin{equation}\label{eq:replicator}
  q_i' = q_i w_i.
\end{equation}
To obtain useful equations of selection, we must consider change. Subtracting $q_i$ from both sides of \Eq{replicator} yields
\begin{equation}\label{eq:aveEx}
  \GD q_i = q_i\left(w_i - 1\right),
\end{equation}
in which $\GD q_i=q_i'-q_i$ is the change in the frequency of each type.

\section{Change caused by selection}

We often want to know about the change caused by selection in the value of a character. Suppose that each type, $i$, has an associated character value, $z_i$.  The average character value in the initial population is $\zbar = \sum q_iz_i$.  The average character value in the descendant population is $\zbar' = \sum q_i'z_i'$.  For now, assume that descendants have the same average character value as their ancestors, $z_i' = z_i$.  Then $\zbar' = \sum q_i'z_i$, and the change in the average value of the character caused by selection is
\begin{equation*}
  \zbar' - \zbar = \GDs \zbar = \sum q_i'z_i - \sum q_iz_i=\sum \left(q_i'-q_i\right)z_i,
\end{equation*}
where $\GDs$ means the change caused by selection when ignoring all other evolutionary forces \autocite{price72fishers,ewens89an-interpretation,frank92fishers}. Using $\GD q_i=q_i'-q_i$ for frequency changes yields
\begin{equation}\label{eq:charChange}
  \GDs \zbar = \sum \GD q_i z_i.
\end{equation}
This equation expresses the fundamental concept of selection \autocite{frank12naturalb}.  Frequencies change according to differences in fitness \Eqp{aveEx}.  Thus, selection is the change in character value caused by differences in fitness, holding constant other evolutionary forces that may alter the character values, $z_i$.  

\section{Change during transmission}

We may consider the other forces that alter characters as the change during transmission. In particular, define $\GD z_i=z_i'-z_i$ as the difference between the average value among descendants derived from ancestral type $i$ and the average value of ancestors of type $i$. Then $\sum q_i'\GD z_i$, is the change during transmission when measured in the context of the descendant population. Here, $q_i'$ is the fraction of the descendant population derived from ancestors of type $i$. 

Thus, the total change, $\GD\zbar=\zbar' - \zbar$, is exactly the sum of the change caused by selection and the change during transmission
\begin{equation}\label{eq:priceEq}
  \GD \zbar = \sum \GD q_i z_i + \sum q_i'\GD z_i,
\end{equation}
a form of the Price equation \autocite{price72extension,frank12naturalb}. We may abbreviate the two components of total change as
\begin{equation}\label{eq:priceEqDelta}
  \GD \zbar = \GDs\zbar + \GDc\zbar,
\end{equation}
which partitions total change into a part ascribed to natural selection, $\GDs$, and a part ascribed to changes in characters during transmission, $\GDc$. The change in transmission subsumes all evolutionary forces beyond selection.

\section{Characters and covariance}

We can express the fundamental equation of selection \Eqp{charChange} in terms of the covariance between fitness and character value. Many of the classic equations of selection derive from the covariance form. Combining Eqs.~(\ref{eq:aveEx}) and (\ref{eq:charChange}) leads to
\begin{equation}\label{eq:dzAveEx}
  \GDs \zbar = \sum \GD q_i z_i = \sum q_i\left(w_i - 1\right)z_i.
\end{equation}
The right-hand side matches the definition for covariance
\begin{equation}\label{eq:charCov}
  \GDs \zbar = \cov(w,z).
\end{equation}
We can rewrite a covariance as a product of a regression coefficient and a variance term
\begin{equation}\label{eq:fitnessReg}
  \GDs\zbar = \cov(w,z) = \Gb_{zw}\var{w},
\end{equation}
where $\Gb_{zw}$ is the regression of phenotype, $z$, on fitness, $w$, and $\var{w}$ is the variance in fitness. The statistical covariance, regression, and variance functions commonly arise in the literature on selection \autocite{robertson66a-mathematical,price70selection,lande83the-measurement,falconer96introduction}. 

\section{Quantitative and genetic characters}

The character $z$ can be a quantitative trait or a gene frequency from the classical equations of population genetics. In a population genetics example, assume that each individual carries one allele. For the $i$th individual, $z_i=0$ when the individual carries the normal allelic type, and $z_i=1$ when the individual carries a variant allele. Then the frequency of the variant allele in the $i$th individual is $p_i=z_i$, the allele frequency in the population is $\bar{p}=\zbar$, and the initial frequencies of each of the $N$ individuals is $q_i=1/N$. From \Eq{dzAveEx}, the change in allele frequency is
\begin{equation}\label{eq:popGenHaploid}
  \GDs \bar{p} = \frac{1}{N} \sum \left(w_i - 1\right)p_i.
\end{equation}
From the prior section, we can write the population genetics form in terms of statistical functions
\begin{equation}\label{eq:popGenPrice}
  \GDs\bar{p} = \cov(w,p) = \Gb_{pw}\var{w}.
\end{equation}

For analyzing allele frequency change, the population genetics form in \Eq{popGenHaploid} is often easier to understand than \Eq{popGenPrice}, which is given in terms of statistical functions. This advantage for the population genetics expression to study allele frequency emphasizes the value of using specialized tools to fit particular problems. 

By contrast, the more abstract statistical form in \Eq{popGenPrice} has advantages when studying the conceptual structure of natural selection and when trying to partition the causes of selection into components. Suppose, for example, that one only wishes to know whether the allele frequency is increasing or decreasing. Then \Eq{popGenPrice} shows that it is sufficient to know whether $\Gb_{pw}$ is positive or negative, because $\var{w}$ is always positive. That sufficient condition is difficult to see in \Eq{popGenHaploid}, but is immediately obvious in \Eq{popGenPrice}.

\section{Variance, distance or information}

The variance in fitness, $\var{w}$, arises in one form or another in every expression of selection. Why is the variance a universal metric of selection? Clearly, variation matters, because selection favors some types over others only when the alternatives differ. But why does selection depend exactly on the variance rather than on some other measure of variation?

I will show that natural selection moves the population a certain distance. That distance is equivalent to the variance in fitness. Thus, we may think about the change caused by selection equivalently in terms of variance or distance.

Begin by noting from \Eq{aveEx} that $\GD q_i/q_i=w_i-1$. Then, the variance in fitness is
\begin{equation}\label{eq:varW}
  \var{w} = \sum q_i(w_i-1)^2=\sum q_i\left(\frac{\GD q_i}{q_i}\right)^2
  	=\sum\frac{\left(\GD q_i\right)^2}{q_i}.
\end{equation}
The squared distance in Euclidean geometry is the sum of the squared changes in each dimension. On the right is the sum of the squares for the change in frequency. Each dimension of squared distance is divided by the original frequency. That normalization makes sense, because a small change relative to a large initial frequency means less than a small change relative to a small initial frequency. The variance in fitness measures the squared distance between the ancestral and descendant population in terms of the frequencies of the types \autocite{ewens92an-optimizing}. 

When the frequency changes are small, the expression on the right equals the Fisher information measure \autocite{frank09natural}. A slightly different measure of information arises in selection equations when the frequency changes are not small \autocite{frank12naturalc}, but the idea is the same. Selection acquires information about environmental challenge through changes in frequency.

Thus, we may think of selection in terms of variance, distance or information. Selection moves the population frequencies a distance that equals the variance in fitness. That distance is equivalent to the gain in information by the population caused by selection.

\section{Characters and coordinates}

We can think of fitness and characters as alternative coordinates in which to measure the changes caused by natural selection in frequency, distance, and information. Using \Eq{aveEx}, we can rewrite the variance in fitness from \Eq{varW} as
\begin{equation*}
  \var{w} = \sum q_i(w_i-1)^2 =\sum\GD q_i w_i.
\end{equation*}
Compare that expression with \Eq{charChange} for the change in character value caused by selection
\begin{equation*}
  \GDs\zbar=\sum\GD q_i z_i.
\end{equation*}
If we start with the right side of the expression for the variance in fitness and then replace $w_i$ by $z_i$, we obtain the change in character value caused by selection. We can think of that replacement as altering the coordinates on which we measure change, from the frequency changes described by fitness, $w_i=q_i'/q_i$, to the character values described by $z_i$.

Although this description in terms of coordinates may seem a bit abstract, it is essential for thinking about evolutionary change in relation to selection. Selection changes frequencies. The consequences of frequency for the change in characters depend on the coordinates that describe the translation between frequency change and characters \autocite{frank12naturalc,frank13natural}. 

Consider the expression from \Eq{priceEq} for total evolutionary change
\begin{equation*}
  \GD \zbar = \sum \GD q_i z_i + \sum q_i'\GD z_i.
\end{equation*}
This is an exact expression that includes four aspects of evolutionary change. First, the change in frequencies, $\GD q_i$, causes evolutionary change. Second, the amount of change depends on the coordinates of characters, $z_i$. Third, the change in the coordinates of characters during transmission, $\GD z_i$, causes evolutionary change. Fourth, the changed coordinates have their consequences in the context of the frequencies in the descendant population, $q_i'$.

\section{Variance and scale}

In models of selection, one often encounters the variance in characters, $\var{z}$, rather than the variance in fitness, $\var{w}$. The variance in characters is simply a change in scale with respect to the variance in fitness---another way in which to describe the translation between coordinates for frequency change and the coordinates for characters. In particular, 
\begin{equation}\label{eq:fitnessReg2}
  \GDs\zbar=\cov(w,z)=\Gb_{wz}\var{w} = \Gb_{zw}\var{z},
\end{equation}
thus $\var{z}=\Gg\var{w}$. Here, $\Gg$ is based on the regression coefficients. The value of $\Gg$ describes the rescaling between the variance in characters and the variance in fitness. Thus, when $\var{z}$ arises in selection equations, it can be thought of as the rescaling of $\var{w}$ in a given context.

\section{Description and causation}

\Eq{fitnessReg2} describes associations between characters and fitness. We may represent those associations as $z \leftrightarrow w$. Here, we know only that a character, $z$, and fitness, $w$, are correlated, as expressed by $\cov(w,z)$. We do not know anything about the causes of that correlation. But we may have a model about how variation in characters causes variation in fitness. To study that causal model, we must analyze how the hypothesized causal structure predicts correlations between characters, fitness and evolutionary change. Alternative causal models provide alternative hypotheses and predictions that can be compared with observation \autocite{crespi90measuring,frank97the-price,frank98foundations,scheiner00using}.

Regression equations provide a simple way in which to express hypothesized causes  \autocite{li75path}. For example, we may have a hypothesis that the character $z$ is a primary cause of fitness, $w$, expressed as a directional path diagram $z\rightarrow w$. That path diagram, in which $z$ is a cause of $w$, is mathematically equivalent to the regression equation
\begin{equation}\label{eq:regi}
  w_i = \phi + \Gb_{wz}z_i+\Ge_i,
\end{equation}
in which $\phi$ is a constant, and $\Ge_i$ is the difference between the actual value of $z_i$ and the value predicted by the model, $\phi+\Gb_{wz}z_i$. 

\section{Partitions of cause}

To analyze causal models, we focus on the general relations between variables rather than on the values of particular types. Thus, we can drop the $i$ subscripts in \Eq{regi} to simplify the expression, as in the following expanded regression equation
\begin{equation}\label{eq:bivarEx}
  w = \phi + \preg{wz}{y}z+ \preg{wy}{z}y+\Ge.
\end{equation}
Here, fitness $w$ depends on the two characters, $z$ and $y$ \autocite{lande83the-measurement}. The partial regression coefficient $\preg{wz}{y}$ is the average effect of $z$ on $w$ holding $y$ constant, and $\preg{wy}{z}$ is the average effect of $y$ on $w$ holding $z$ constant. Regression coefficients minimize the total distance (sum of squares) between the actual and predicted values. Minimizing the residual distance maximizes the use of the information contained in the predictors about the actual values.

This regression equation is exact, in the sense that it is an equality under all circumstances. No assumptions are needed about additivity or linearity of $z$ and $y$ or about normal distributions for variation. Those assumptions arise in statistical tests of significance when comparing the regression coefficients to hypothesized values or when predicting how the values of the regression coefficients change with context. 

Note that the regression coefficients, $\Gb$, often change as the values of $w$ or $z$ or $y$ change, or if we add another predictor variable. The exact equation is a description of the relations between the variables as they are given. The structure of the relations between the variables forms a causal hypothesis that leads to predictions \autocite{li75path}.

\begin{figure*}[t]
\centering
\includegraphics[width=4in]{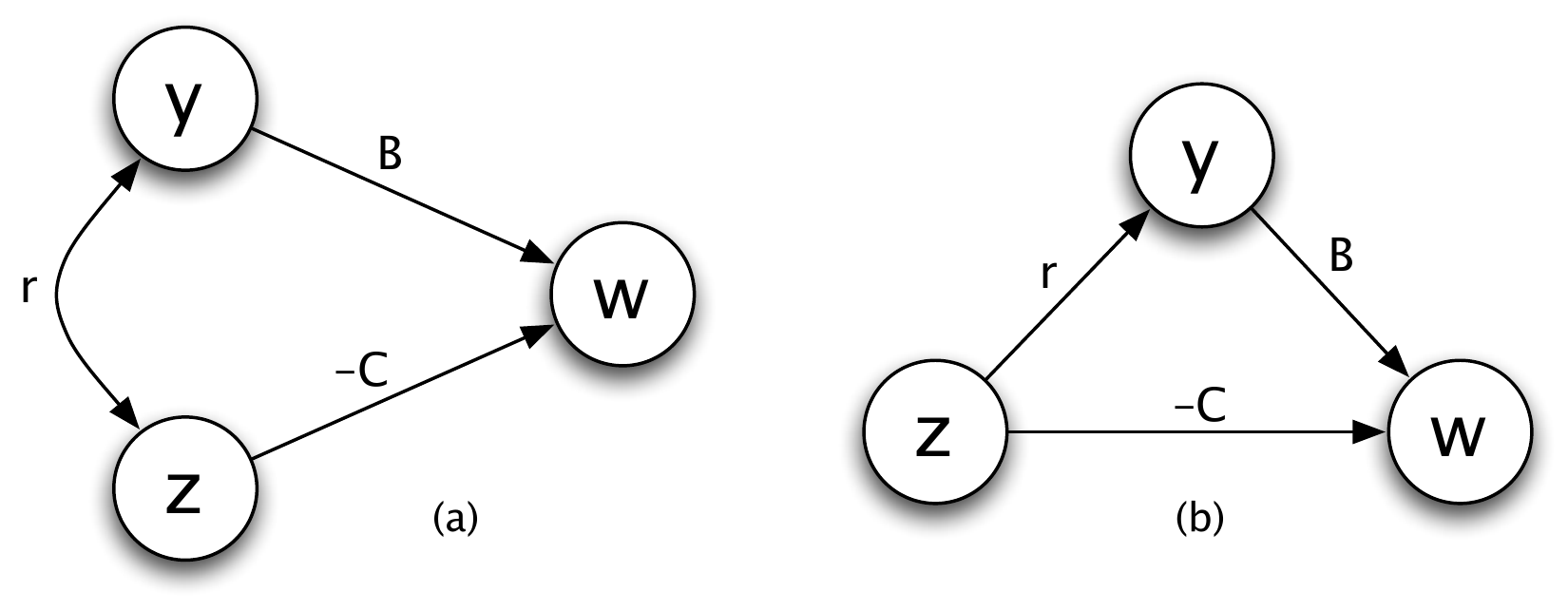}
\caption{Path diagrams for the effects of phenotype, $z$, and secondary predictor, $y$, on fitness, $w$.  (a) An unknown cause associates $y$ and $z$. The arrow connecting those factors points both ways, indicating no particular directionality in the hypothesized causal scheme.  (b) The phenotype, $z$, directly affects the other predictor, $y$, which in turn affects fitness.  The arrow pointing from $z$ to $y$ indicates the hypothesized direction of causality. The choice of notation matches kin selection theory, in which $z$ is an altruistic behavior that reduces the fitness of an actor by the cost $C$ and aids the fitness of a recipient by the benefit, $B$, and $r$ measures the association between the behaviors of the actor and recipient. Although that notation comes from kin selection theory, the general causal scheme applies to any pair of correlated characters that influences fitness \autocite{lande83the-measurement,queller92a-general}. From \textcite{frank13natural}.}
\label{fig:fitPath}
\end{figure*}

\section{Partitions of fitness}

We can interpret \Eq{bivarEx} as a hypothesis that partitions fitness into two causes. Suppose, for example, that we are interested in the direct effect of the character $z$ on fitness. To isolate the direct effect of $z$, it is useful to consider how a second character, $y$, also influences fitness (\Fig{fitPath}). 

The condition for $z$ to increase by selection can be evaluated with \Eq{fitnessReg2}. That equation simply states that $z$ increases when it is positively associated with fitness.  However, we now have the complication shown in \Eq{bivarEx} that fitness also depends on another character, $y$.  If we expand $\cov(w,z)$ in \Eq{fitnessReg2} with the full expression for fitness in \Eq{bivarEx}, we obtain
\begin{equation}\label{eq:dzHR}
  \GDs\zbar = \left(\preg{wz}{y} + \preg{wy}{z} \Gb_{yz}\right)\var{z}.
\end{equation}
Following \textcite{queller92a-general}, I abbreviate the three regression terms. The term, $\Gb_{yz}=r$, describes the association between the phenotype, $z$, and the other predictor of fitness, $y$. An increase in $z$ by the amount $\GD z$ corresponds to an average increase of $y$ by the amount $\GD y = r\GD z$. The term, $\preg{wy}{z}=B$, describes the direct effect of the other predictor, $y$, on fitness, holding constant the focal phenotype, $z$.  The term, $\preg{wz}{y}=-C$, describes the direct effect of the phenotype, $z$, on fitness, $w$, holding constant the effect of the other predictor, $y$.  

The condition for the increase of $z$ by selection is $\GDs\zbar > 0$.  The same condition using the terms on the right side of \Eq{dzHR} and the abbreviated notation is
\begin{equation}\label{eq:rbc}
  rB - C > 0.
\end{equation}
This condition applies whether the association between $z$ and $y$ arises from some unknown extrinsic cause (\Fig{fitPath}a) or by the direct relation of $z$ to $y$ (\Fig{fitPath}b). 

This expression in \Eq{rbc} describes the condition for selection to increase the character, $z$, when ignoring any changes in the character that arise during transmission. Thus, when one wants to know whether selection acting by this particular causal scheme would increase a character, it is sufficient to know if this simple condition holds.

\section{Testing causal hypotheses}

If selection favors an increase in the character $z$, then the condition in \Eq{rbc} will always be true. That condition simply expresses the fact that the slope of fitness on character value, $\Gb_{wz}$, must be positive when selection favors an increase in $z$. The expression $rB-C$ is one way in which to partition $\Gb_{wz}$ into components. However, the fact that $rB-C>0$ does not mean that the decomposition into those three components provides a good causal explanation for how selection acts on the character $z$. 

There are many alternative ways in which to partition the total effect of selection into components. Other characters may be important. Environmental or other extrinsic factors may dominate. How can we tell if a particular causal scheme is a good explanation? 

If we can manipulate the effects $r$, $B$ or $C$ directly, we can run an experiment. If we can find natural comparisons in which those terms vary, we can test comparative hypotheses. If we add other potential causes to our model, and the original terms hold their values in the context of the changed model, that stability of effects under different conditions increases the likelihood that the effects are true.

Three points emerge. First, a partition such as $rB-C$ is sufficient to describe the direction of change, because a partition simply splits the total change into parts. Second, a partition does not necessarily describe causal relations in an accurate or useful way. Third, various methods can be used to test whether a causal hypothesis is a good explanation.

\section{Multiple characters and nonadditivity}

Instead of the simple causal schemes in \Fig{fitPath}, there may be multiple characters $y_j$ correlated with $z$. Then the condition for the increase of $z$ becomes $\sum r_jB_j - C > 0$, in which $r_j$ is the regression of $y_j$ on $z$, and $B_j$ is the partial regression of $w$ on $y_j$, holding constant all other characters. This method also applies to multiplicative interactions between characters. For example, suppose $\pi_{12}=y_1\times y_2$, for characters $y_1$ and $y_2$. Then we can use $\pi_{12}$ as a character in the same type of analysis, in which $r$ would be the regression of $\pi_{12}$ on $z$, and $B$ would be the partial regression of $w$ on $\pi_{12}$ holding constant all other variables.

\section{Partitions of characters}

We have been studying the partition of fitness into separate causes, including the role of individual characters. Each character may itself be influenced by various causes. Describe the cause of a character by a regression equation
\begin{equation*}
  z = \phi+\Gb_{zg}g + \Gd,
\end{equation*}
in which $\phi$ is a constant that is traditionally set to zero in this equation, $g$ is a predictor of phenotype, the regression coefficient $\Gb_{zg}$ is the average effect of $g$ on phenotype $z$, and $\Gd=z-\Gb_{zg}g$ is the residual between the actual value and the predicted value. This regression expression describes phenotypic value, $z$, based on any predictor, $g$.  For predictors, we could use temperature, neighbors' behavior, another phenotype, epistatic interactions given as the product of allelic values, symbiont characters, or an individual's own genes.  

\textcite{fisher18the-correlation} first presented this regression for phenotype in terms of alleles as the predictors. Suppose 
\begin{equation}\label{eq:gDef}
  g = \sum_j b_j x_j,
\end{equation}
in which $x_j$ is the presence or absence of an allelic type. Then each $b_j$ is the partial regression of an allele on phenotype, which describes the average contribution to phenotype for adding or subtracting the associated allelic type. The coefficient $b_j$ is called the average allelic effect, and $g$ is called the breeding value \autocite{fisher30the-genetical,crow70an-introduction,falconer96introduction}.  When $g$ is defined as the sum of the average effects of the underlying predictors, then $\Gb_{zg}=1$, and 
\begin{equation}\label{eq:zReg}
  z = g + \Gd,
\end{equation}
where $\Gd = z-g$ is the difference between the actual value and the predicted value.  

Some facts will be useful in the next section. If we take the average of both sides of \Eq{zReg}, we get $\zbar = \gbar$, because $\bar{\Gd}=0$ by the theory of regression. If we take the variance of both sides, we obtain $\var{z}=\var{g}+\var{\Gd},$ noting that, by the theory of regression, $g$ and $\Gd$ are uncorrelated.

\section{Heritability and the response to selection}

To study selection, we first need an explicit form for the relation between character value and fitness, which we write here as
\begin{equation*}
  w = \phi+\Gb_{wz}z + \Ge.
\end{equation*}
Substitute that expression into the covariance expression of selection in \Eq{fitnessReg2}, yielding
\begin{equation}\label{eq:sV}
  \GDs\zbar=\cov(w,z)=\Gb_{wz}\var{z} = s\var{z},
\end{equation}
because $\phi$ is a constant and $\Ge$ is uncorrelated with $z$, causing those terms to drop out of the covariance. Here, the selective coefficient $s=\Gb_{wz}$ is the effect of the character on fitness. Expand $s\var{z}$ by the partition of the character variance given in the previous section, which leads to
\begin{equation}\label{eq:partSel}
  \GDs\zbar=s\var{z}=s\var{g}+s\var{\Gd} = \GDg\zbar + \GDn\zbar.
\end{equation}
We can think of $g$ as the average effect of the predictors of phenotype that we have included in our causal model of character values.  Then $s\var{g}=\GDg\zbar$ is the component of total selective change associated with our predictors, and
\begin{equation}\label{eq:partSelG}
  \GDg\zbar=\GDs\zbar - \GDn\zbar,
\end{equation}
shows that the component of selection transmitted to descendants through the predictors included in our model, $\GDg$, is the change caused by selection, $\GDs$, minus the part of the selective change that is not transmitted through the predictors, $\GDn$. Although it is traditional to use alleles as predictors, we can use any hypothesized causal scheme. Thus, the separation between transmitted and nontransmitted components of selection depends on the hypothesis for the causes of phenotype.

If we choose the predictors for $g$ to be the individual alleles that influence phenotype, then $\var{g}$ is the traditional measure of genetic variance, and $s\var{g}$ is that component of selective change that is transmitted from parent to offspring through the effects of the individual alleles. The fraction of the total change that is transmitted, $\var{g}/\var{z}$, is a common measure of heritability. 

\section{Changes in transmission and total change}

This section describes the total evolutionary change when considered in terms of the parts of phenotype that are transmitted to descendants. Here, the transmitted part arises from the predictors in an explicit causal hypothesis about phenotype. 

The expression of characters in terms of predictors from \Eq{zReg} is $z=g+\Gd$. From that equation, $\zbar=\gbar$, because the average residuals of a regression, $\bar{\Gd}$, are zero. Thus, when studying the change in a character, we have $\GD\zbar=\GD\gbar$, which means that we can analyze the change in a character by studying the change in the average effects of the predictors of a character. Thus, from \Eq{priceEq}, we may write the total change in terms of the coordinates of the average effects of the predictors, $g$, yielding
\begin{equation}\label{eq:deltaGT}
  \GD\zbar = \sum\GD q_i g_i + \sum q_i'\GD g_i= \GDg\zbar + \GDt\zbar,
\end{equation}
in which $\GDt\zbar$ is the change in the average effects of the predictors during transmission \autocite{frank97the-price,frank98foundations}. The total change divides into two components: the change caused by the part of selection that is transmitted to descendants plus the change in the transmitted part of phenotype between ancestors and descendants. Alternatively, we may write $\GDg\zbar =\GDs\gbar$, the total selective component expressed in the coordinates of the average effects of the predictors, and $\GDt\zbar=\GDc\gbar$, the total change in coordinates with respect to the average effects of the predictors.

\section{Choice of predictors}

If natural selection dominates other evolutionary forces, then we can use the theory of natural selection to analyze evolutionary change. When does selection dominate? From \Eq{deltaGT}, the change in phenotype caused by selection is $\GDg$. If the second term $\GDt$ is relatively small, then we can understand evolutionary change primarily through models of selection. 

A small value of the transmission term, $\GDt$, arises if the effects of the predictors in our causal model of phenotype remain relatively stable between ancestors and descendants. Many factors may influence phenotype, including alleles and their interactions, maternal effects, various epigenetic processes, changing environment, and so on. Finding a good causal model of phenotype in terms of predictors is an empirical problem that can be studied by testing alternative causal schemes against observation. 

Note that the equations of evolutionary change do not distinguish between different kinds of predictors. For example, one can use both alleles and weather as predictors. If weather varies among types and its average effect on phenotype transmits stably between ancestors and descendants, then weather provides a useful predictor. Variance in stably transmitted weather attributes can lead to changes in characters by selection. 

Calling the association between weather and fitness an aspect of selection may seem strange or misleading. One can certainly choose to use a different description. But the equations themselves do not distinguish between different causes.

\section{Sufficiency and invariance}

To analyze natural selection, what do we need to know? Let us compare two alternatives. One provides full information about how the population evolves over time. The other considers only how natural selection alters average character values at any instant in time. 

A full analysis begins with the change in frequency given in \Eq{aveEx}, as
\begin{equation*}
  \GD q_i = q_i\left(w_i - 1\right).
\end{equation*}
For each type in the population, we must know the initial frequency, $q_i$, and the fitness, $w_i$. From those values, each new frequency can be calculated. Then new values of fitnesses  would be needed to calculate the next round of updated frequencies. Fitnesses can change with frequencies and with extrinsic conditions. That process provides a full description of the population dynamics over time. The detailed output concerning dynamics reflects the detailed input about all of the initial frequencies and all of the fitnesses over time.

A more limited analysis arises from the part of total evolutionary change caused by selection. If we focus on the change by selection in the average value of a character at any point in time, we have
\begin{equation*}
  \GDs\zbar = \sum\GD q_iz_i = \cov(w,z)=\Gb_{zw}\var{w},
\end{equation*}
from Eqs.~(\ref{eq:dzAveEx}) and (\ref{eq:fitnessReg}). To calculate the average change caused by selection, it is sufficient to know the covariance between the fitnesses and character values over the population. We do not need to know the individual frequencies or the individual fitnesses. A single summary statistic over the population is sufficient. A single assumed input corresponds to a single output. We could, of course, make more complicated assumptions and get more complicated outputs. What we get out matches what we put in.

Invariance provides another way to describe sufficiency. The mean change in character value caused by selection is invariant to all aspects of variability except the covariance. The reason is that the variance in fitness, $\var{w}$, describes the distance the population moves with regard to frequencies, and the regression $\Gb_{zw}$ rescales the distance along coordinates of frequency into distance along coordinates of the character. 

The analysis of two characters influencing fitness provides another example of sufficiency and invariance. The condition for the average value of the focal character to increase by selection is given by $rB-C>0$ in \Eq{rbc}. That condition shows that all other details about variability and correlation between the two characters and fitness do not matter. Thus, the direction of change caused by selection is invariant to most details of the population. Simple invariances of this kind often provide great insight into otherwise complex problems \autocite{frank13naturalvii}.

Fisher's fundamental theorem of natural selection is a simple invariance \autocite{frank12wrights}. The theorem states that, at any instant in time, the change in average fitness caused by selection is equal to the variance in fitness. Fisher was particularly interested in the transmissible component of fitness based on a causal model of alleles. Thus, his variance in fitness is $\var{g}$ from \Eq{partSel} based on the average effects of alleles on fitness. Fisher's theorem shows that the change in mean fitness by selection is invariant to all details of variability in the population except the variance associated with the transmissible predictors.

\section{Transmission versus selection}

\begin{quote}
In evolutionary theory, a gene could be defined as any hereditary information for which there is a $\ldots$ selection bias equal to several or many times its rate of endogenous change \autocite{williams66adaptation}.
\end{quote}

Selection and transmission often oppose each other. Selection increases fitness; mutation decays fitness during transmission. Selection among groups favors cooperation; selection within groups favors selfishness that decays the transmission of cooperative behavior.

Total change in terms of selection and transmission \Eqp{priceEqDelta} is
\begin{equation*}
  \GD\zbar=\GDs\zbar+\GDc\zbar=\GDS + \GDT,
\end{equation*}
which may alternatively be expressed in terms of predictors, as in \Eq{deltaGT}. An equilibrium balance between selection and mutation, or between different levels of selection, occurs when
\begin{equation}\label{eq:seltrEq}
  \GDS=-\GDT.
\end{equation}
The strength of selection bias relative to endogenous change during transmission is
\begin{equation}\label{eq:selTrRatio}
  \R=\log\left|\frac{\GD S}{\GD\Gt}\right|,
\end{equation} 
assuming that the forces oppose \autocite{frank12natural}. The logarithm provides a natural measure of relative strength, centered at zero when the balance in \Eq{seltrEq} holds.

We may write the selection term as $\GDS=s\var{z}$ from \Eq{fitnessReg2}, in which the selective coefficient $s=\Gb_{wz}$ is the slope of fitness on character value. Then, if the equilibrium in \Eq{seltrEq} exists, the character variance is 
\begin{equation}\label{eq:vzEq}
  \var{z}=\frac{-\GDT}{s}.
\end{equation}

\begin{figure*}[!t]
\centering
\includegraphics[width=5in]{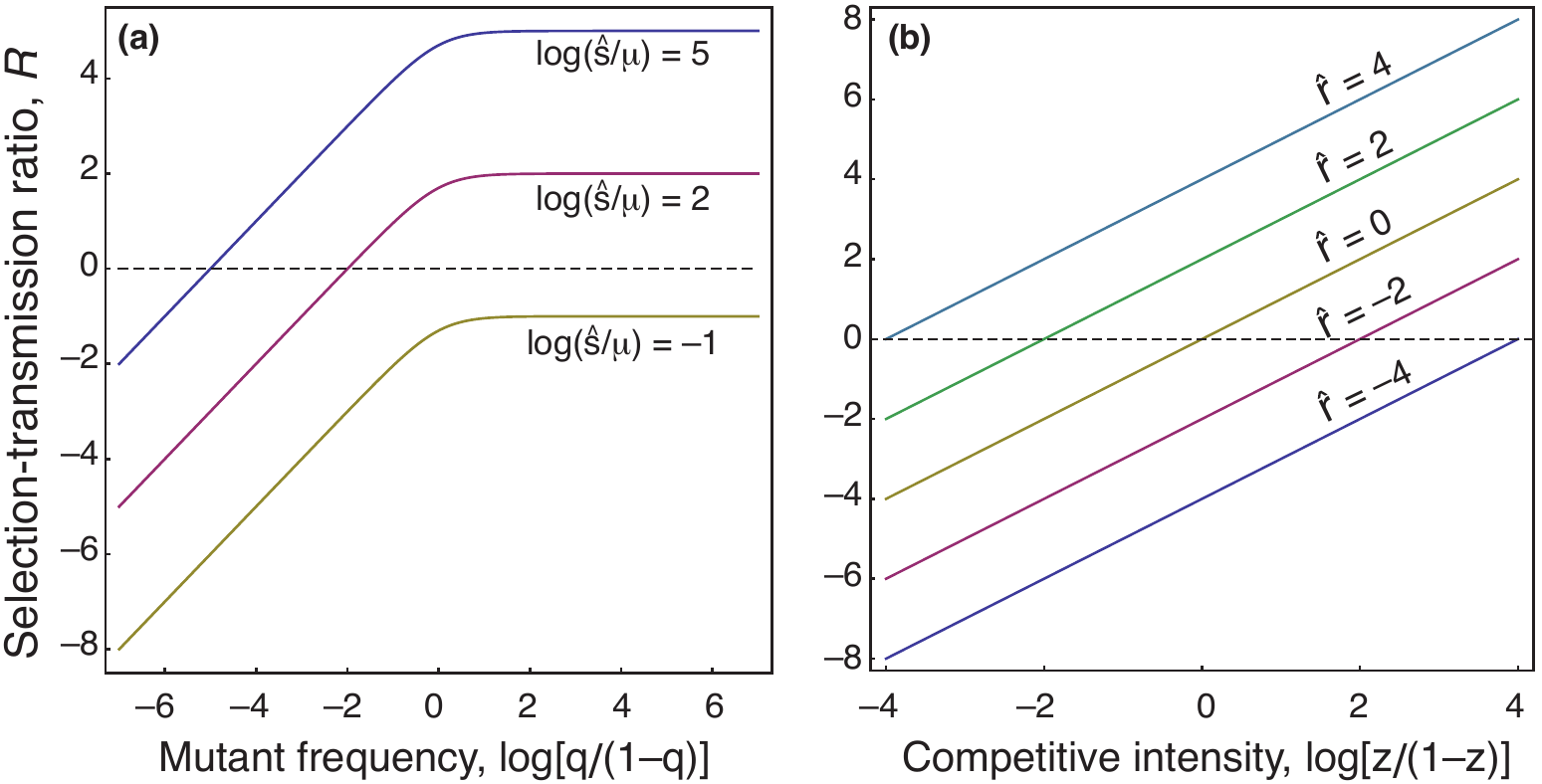}
\caption{The opposing forces of selection and transmission change with context. Equilibrium occurs when the forces balance at $\R=0$ in \Eq{selTrRatio}. In these examples, when the ratio $\R$ is positive, selection dominates and pushes down the character values, and when $\R$ is negative, transmission bias dominates and pushes up character values. (a) The opposition of selection and mutation \Eqp{logRatiomutsel}. (b) The opposition of selection bias among groups versus transmission bias within groups \Eqp{balanceGroup}, with $\hat{r}=r/(1-r)$. All logarithms use base 10. From \textcite{frank12natural}.}
\label{fig:logRatio}
\end{figure*}

\section{Mutation versus selection}

Suppose each individual has one allele. Let a normal allele have character value $z=0$ and a mutant allele have value $z=1$. Then the average character value in the population, $\zbar\equiv q$, is the frequency of the mutant allele. The variance in the character is the binomial variance of the allele frequency, $\var{z}=q(1-q)$, thus $\GDS=sq(1-q)$. The selective intensity against the mutant allele, $s$, is negative, because the mutant decreases fitness. We may use $\hat{s}=-s$ to obtain a positive value for the intensity of selection. The change in frequency during transmission is $\GDT=\Gm(1-q)$, in which normal alleles at frequency $1-q$ are changed into mutant alleles at a rate $\Gm$. 

Using these expressions in the balance between selection bias and transmission bias \Eqp{seltrEq} yields the equilibrium frequency of the mutant allele
\begin{equation}\label{eq:mutSelq}
  q=\frac{\Gm}{\hat{s}}.
\end{equation}
The ratio of selective intensity to transmission bias is
\begin{equation}\label{eq:logRatiomutsel}
  \R=\log\left|\frac{\hat{s}q}{\Gm}\right|.
\end{equation}
This ratio depends on frequency, $q$. When $\hat{s}q>\Gm$, selection pushes down the mutant frequency. When $\hat{s}q<\Gm$, mutation pushes up the mutant frequency. The relative strength of selection to transmission is often frequency dependent (\Fig{logRatio}a). 

\section{Stabilizing selection}

The previous example concerned directional selection. In that case, mutation adds deleterious mutations and selection removes those mutations. Alternatively, stabilizing selection may favor an optimal character value, and mutation may tend to move the character value away from the optimum. 

For example, fitness may drop off with the squared distance from the optimum. Let the underlying character value be $y$, and the squared distance from the optimum $y^*$ be $z=(y-y^*)^2$. The drop off in fitness with squared deviation from the optimum is given by $s=\Gb_{wz}=-\hat{s}$.

If mutations occur with probability $\Gm$, and a mutation has an equal chance of increasing or decreasing character value by $c$, then the transmission bias in terms of squared deviations is $\GDT=\Gm c^2=\var{\Gm}$, in which we can think of $\GDT$ as the variance added by mutation, $\var{\Gm}$. From \Eq{vzEq}, the equilibrium balance between mutation and selection occurs at
\begin{equation*}
  \var{z}=\frac{\var{\Gm}}{\hat{s}}.
\end{equation*}

\section{Clade selection}

The opposition of selection at different levels is similar to the balance between mutation and selection. For example, \textcite{van-valen75group} argued that sexuality may increase the reproductive rate of clades by enhancing the speciation rate. That advantage in competition between clades may be offset by the disadvantage of sexuality within clades, because sexual reproduction is less efficient than asexual reproduction. 

If the selection bias between clades favoring sexuality is $\hat{s}$, and the transmission bias against sexuality within clades is $\Gm$, then the approximate equilibrium frequency $q$ of asexuality would be
\begin{equation*}
  q\approx\frac{\Gm}{\hat{s}},
\end{equation*}
matching the simple genetic model of mutation-selection balance in \Eq{mutSelq}. 

Van Valen also applied this approach to mammals. In mammals, genera with larger body size survive longer than genera with smaller body size, but the smaller-bodied genera produce new genera at a higher rate. The net reproductive rate of small genera is higher, giving a selective advantage to small-bodied genera over large-bodied genera. Within genera, there is a bias towards larger body size. The distribution of mammalian body size is influenced by the balance between selection among genera favoring smaller size and selection within genera favoring larger size.

\section{Kin and group selection}

Van Valen considered clade selection as roughly similar to the balance of selection among clades and mutation within clades. A more accurate description arises by considering the relative strength of selection at the higher and lower levels \textcite{frank12natural}. 

Start with the equality $\GDS=-\GDT$ that describes the fundamental balance between selection bias and transmission bias \Eqp{seltrEq}. We may rewrite the balance as $\GDS\ssA=-\GDS\ssW$, the equality of selection among groups and selection within groups. We can express selective change as $\GDS=s\var{}$, the product of selective intensity and character variance \Eqp{sV}. Thus, the balance between selection at a higher level and transmission bias caused by selection at a lower level is
\begin{equation*}
  s\ssA\varss{A}=-s\ssW\varss{W}.
\end{equation*}
The total variance is the sum of the variances among and within groups, $\varss{T}=\varss{A}+\varss{W}$, thus we may write
\begin{equation*}
  s\ssA\varss{A}=-s\ssW(\varss{T}-\varss{A}).
\end{equation*}
Dividing both sides by the total variance, $\varss{T}$, yields
\begin{equation}\label{eq:balanceR}
  s\ssA r=-s\ssW(1-r),
\end{equation}
in which $r=\varss{A}/\varss{T}$ measures the correlation in character values between individuals within groups or, equivalently, the ratio of the variance among groups to the total variance. 

In general, the variance $\var{}$ provides a weighting that describes the consequences of selection. Thus, $r$ can be thought of as the fraction of the total weighting of selection that happens at the group level, and $1-r$ can be thought of as the fraction of the total weighting of selection that happens within groups. Here, $r$ may often be interpreted as a form of the regression coefficient of relatedness from kin selection theory \autocite{hamilton70selfish}. Thus, we may think of this balance between levels of selection in terms of either kin or group selection \autocite{hamilton75innate,frank86hierarchical}.

\section{Competition versus cooperation}

This section illustrates the balance between opposing selection at different levels. Suppose fitness is
\begin{equation}\label{eq:tragedyW}
  w=\frac{z}{z\ssW}(1-z\ssW).
\end{equation}
The first term describes the competitiveness of a type, $z$, relative to the average competitiveness of its neighbors within a group, $z\ssW$. The second term, $1-z\ssW$, describes the success of the group in competition against other groups \autocite{frank94kin-selection,frank95mutual}.

We can use \Eq{balanceR} to evaluate the balance between selection at the group level and selection within groups. The selective intensity among groups is $s\ssA=-1$, the slope of group fitness, $1-z\ssW$, relative to group phenotype, $z\ssW$.  The selective intensity within groups is $s\ssW=(1-z\ssW)/z\ssW$, which is the change in individual fitness, $w$, with the change in individual character value, $z$, holding constant group phenotype, $z\ssW$.

Substituting these values for $s\ssA$ and $s\ssW$ into \Eq{balanceR} yields a balance, $\GDS=-\GDT$, between group and individual selection as
\begin{equation}\label{eq:balanceGroup}
  -r=-\frac{1-z\ssW}{z\ssW}(1-r).
\end{equation}
If the population approaches an equilibrium balance between the opposing components of selection, then individual and group phenotypes converge such that $z=z\ssW=z^*$, and we obtain
\begin{equation*}
  z^*=1-r.
\end{equation*}
The coefficient $r$ measures the similarity of individuals within groups or, equivalently, the variance in phenotype among groups. As $r$ increases, competitiveness $z^*$ declines. In this model, reduced competitiveness may be thought of as increased cooperation. Thus, $r$ determines the balance between cooperation favored by selection among groups and competitiveness favored by the transmission bias of selection within groups (\Fig{logRatio}b).

\section{ESS by fitness maximization}

We could partition the fitness expression in \Eq{tragedyW} into components by using regression equations. That approach would split fitness into the part explained by the character for individual competitiveness, $z$, and the part explained by group competitiveness, $z\ssW$. Each part would be weighted by its partial regression of fitness on the character, holding the other character constant. The problem is that the nonlinearity in \Eq{tragedyW} makes it difficult to calculate the regression coefficients. 

If we assume that character variances are small, the following trick often makes the analysis very easy \autocite{frank95mutual,frank97multivariate,frank98foundations,taylor96how-to-make}. With small variances, the partial regression coefficients of $w$ with respect to each character will be close to the partial derivative of $w$ with respect to each character. For two characters, $z$ and $y$, the condition for selection to favor an increase in $z$ is
\begin{equation}\label{eq:partRBC}
  \dovr{w}{z} = \povr{w}{y}\dovr{y}{z} + \povr{w}{z} > 0.
\end{equation}
Because variances are small, we can relate each derivative term to a regression coefficient and also to the $r$, $B$ and $C$ terms of \Fig{fitPath}. In particular, $\dd w/\dd z = \Gb_{wz}$, describes the slope of fitness on character value, which determines the direction of selection. That total change depends on $\dd y/\dd z = \Gb_{yz}=r$, which describes the slope of the character $y$ relative to the focal character $z$. The term $\prt w/\prt y=\preg{wy}{z}=B$ describes the slope of fitness on $y$, holding constant $z$.  The term, $\prt w/\prt z=\preg{wz}{y}=-C$, describes the direct effect of $z$ on fitness, holding constant, $y$. 

Using those definitions, the condition reduces to $rB-C>0$. The method extends to any number of characters. The advantage here is that the specific terms follow automatically from differentiation, given any expression for the functional relation between fitness and various characters. The method assumes that character variances are small and that characters vary continuously. The approach works well in finding a potential equilibrium that is an evolutionary stable strategy (ESS), which means that the equilibrium character value tends to outcompete other character values that differ by a small amount \autocite{maynard-smith73the-logic,maynard-smith82evolution}. 

For simple problems, the equilibrium reduces to $rB=C$, which balances opposing components of selection. If we use the fitness expression in \Eq{tragedyW}, letting $y\equiv z\ssW$, and assume that variances are small so that near a balance we have $z=z\ssW=z^*$, then the method in \Eq{partRBC} allows the use of simple differentiation to obtain $z^*=1-r$.

\section{Dynamics of phenotypic evolution}

The differentiation method simplifies study of the equilibrium balance between opposing components of selection \Eqp{partRBC}. We only need an expression for the causal relations between fitness and characters. The main problem concerns dynamics. Real populations may never follow a path to the equilibrium balance of forces. To study dynamics, one could specify all aspects of how genes and other predictors affect phenotypes, and all aspects of spatial and temporal variability. Then one could calculate the dynamical path along which populations evolve.  

Equilibrium balance and dynamical analysis are two distinct tools, each with advantages and limitations. The equilibrium models require few specific assumptions. They describe simply how changed assumptions alter predicted outcomes. But simple equilibrium models ignore many constraints and other forces that must be present in any real case. A mathematician trained in the careful analysis of dynamical models often finds such extreme simplifications abhorrent. So many things are either unspecified or potentially wrong, that almost certainly some aspect of the analysis will be misleading or incorrect.

The dynamical models require more assumptions and provide more insight into the evolutionary paths that populations may follow. But those models, by requiring significant detail as input, may be exact mathematical analyses that apply exactly to nothing. Biologists trying to understand natural history are often puzzled by the use of so many precise assumptions that cannot match any real aspect of biology.

Each approach, applied judiciously, reflects a different aspect of reality. Given the complexities of most biological problems, the simpler equilibrium method has been much more successful in the design and interpretation of empirical studies. The more detailed dynamical models help most when exploring, in theory, potentially complex interactions that would be difficult study by simpler approaches.

\section{Other topics}

Evolutionary conflict arises when the causes of characters transmit in different ways. Different transmission pathways lead to opposing relations between character values and fitness. Simple theories of selection identify such conflicts. The complexity of conflict creates challenges in analyzing the causes of selection \autocite{burt08genes}.

Drifting frequencies arise in small populations. When relatively few individuals reproduce, random processes often influence success more strongly than differences in character values. The evolution of characters may depend on chance events rather than adaptive changes favored by selection \autocite{crow70an-introduction}.

Variable selection arises when environments change over time or space. A character with lower average success and lower variability in success may be favored relative to a character with higher average success and higher variability. The theory closely matches economic concepts of risk aversion \autocite{gillespie73natural,frank11natural}.

Reproductive valuation arises when individuals in different classes contribute differently to future populations. Old individuals may have less chance of future reproduction than young individuals. When evaluating the potential success of a character, one must consider if the character occurs differently between classes of individuals \autocite{charlesworth94evolution}. 

Nonheritable variation arises when the same set of genes or predictors leads to diverse character values. Fitness of a type is the average over the range of character values expressed, altering the relation between transmissible predictors of characters and fitness. Variability in expression can itself be a character influenced by selection \autocite{west-eberhard03developmental,frank11naturalb}.

Evolvability arises when a character increases the potential for future evolution. A character may raise the chance of producing novel phenotypes. Future potential for exploring phenotypic novelty may reduce immediate fitness. Selection theory must be extended to evaluate conflicting forces at different timescales \autocite{wagner96complex}.

\section{Deductive versus inductive perspectives}

Sometimes it makes sense to think in terms of deductive predictions. What do particular assumptions about initial conditions, genetic interactions, fitness of individuals, and spatial complexity predict about evolutionary dynamics? 

Sometimes it makes sense to think in terms of inductive analysis. Given certain frequency changes and the total distance between ancestor and descendant populations, how much do different causes explain of that total distance?

Mathematical theories often analyze deductive models of dynamics. Practical applications to empirical problems often inductively partition causes. In practical applications, one asks: How well do various alternative causal structures fit with the observed or assumed pattern of change? What character values are causally consistent with lack of change near an equilibrium?

The deductive and inductive approaches each have benefits. Deductive approaches often provide the only way to study the consequences of particular assumptions. Inductive approaches often provide the only way to analyze the causes of particular patterns. This article emphasized the inductive analysis of cause. 

Consider random drift and selection. Deductively, one assumes randomness in small populations and differences in expected reproductive success. From those assumptions, one calculates the probability that a population ends up in a particular state. 

Inductively, one starts with an observed or assumed total distance between the initial and final population. Causal hypotheses partition that total change into random and selective components---one component caused by random sampling processes and one component caused by characters that influence reproductive success. Which of the alternative causal partitions of total change best fits all of the assumptions or all of the available data?

The nature of selection encourages an inductive perspective \autocite{frank09natural,frank12naturalb}. Populations change in frequency composition. Those frequency changes---the actual distance between the ancestral and descendant populations---cause populations to acquire information inductively about the environment. The transmissible predictors of characters correlated with fitness determine the fraction of the inductively acquired information retained by the population.

\section*{Acknowledgments}

National Science Foundation grant DEB--1251035 supports my research. Parts of this chapter were taken from my series of articles on natural selection published in the Journal of Evolutionary Biology.

\bigskip
\bibliography{main}

%
%

\end{document}